\documentclass[cameraready]{Interspeech}

\title{Exploiting Neural Audio Codec Latents for Adversarial Audio Attacks}

\author[equalcontribution]{Sameek}{Bhattacharya}
\author[equalcontribution]{Bharath}{Krishnamurthy}
\author[correspondingauthor]{Ajita}{Rattani}

\address{
    Dept. of Computer Science and Engineering, \\University of North Texas, Denton, Texas, USA
}

\email{sameekbhattacharya@my.unt.edu, bharathkrishnamurthy@my.unt.edu, ajita.rattani@unt.edu}


\usepackage{comment}
\usepackage{booktabs}
\usepackage{amsmath}
\usepackage{amssymb}
\usepackage{pifont}
\usepackage{multirow}
\usepackage{subcaption}
\usepackage{balance}
\usepackage{cuted}
\usepackage{threeparttable}

\usepackage{booktabs}
\usepackage{siunitx}

\begin{document}
\maketitle

\begin{figure*}
    \centering
    \includegraphics[width=0.90\linewidth]{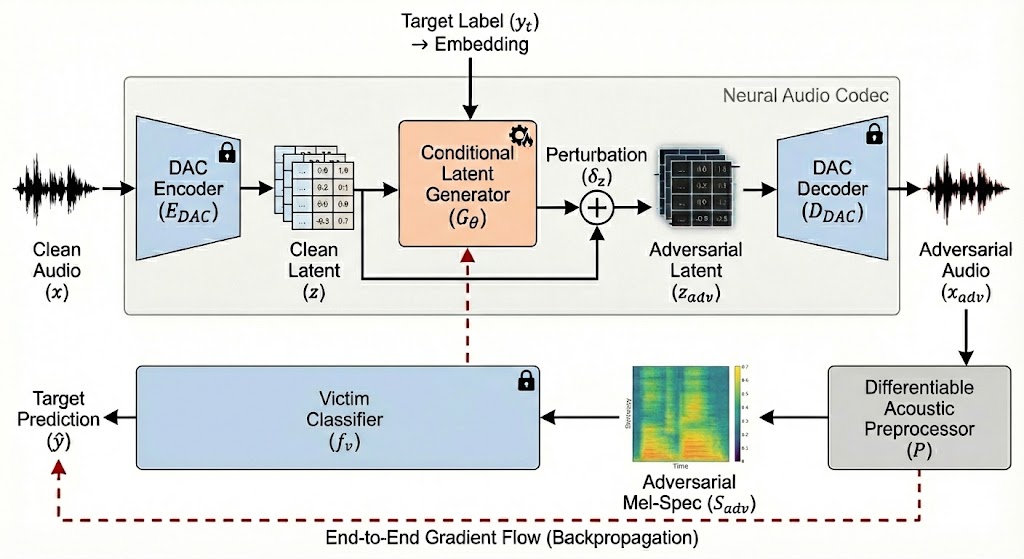}
    \caption{Architectural overview of the proposed end-to-end latent-space adversarial attack framework. The clean input waveform $x$ is first projected into a continuous, lower-dimensional manifold $z$ via the frozen DAC Encoder ($E_{DAC}$). The trainable conditional generator ($G_\theta$, highlighted in orange) synthesizes a targeted perturbation $\delta_z$ by fusing the pristine latent $z$ with the target class embedding $y_t$. The perturbed latent $z_{adv}$ is then reconstructed into the adversarial waveform $x_{adv}$ through the frozen DAC Decoder ($D_{DAC}$). To bridge the time-domain audio and the frozen victim classifier, a fully differentiable acoustic feature extractor ($\mathcal{P}$) is utilized. The dashed line illustrates the continuous gradient flow (backpropagation) from the victim classifier's composite loss all the way back to the latent generator, enabling single-shot, waveform-optimized attacks without relying on iterative inference-time optimization.}
    \label{fig:pipeline}
\end{figure*}

\begin{abstract}

Deep learning–based audio classification systems, including automatic speaker verification, are vulnerable to adversarial attacks. Realistic real-time threat assessment remains difficult because optimization-based methods, such as projected gradient descent (PGD) and Carlini–Wagner, require costly iterative updates in the high-dimensional waveform domain. Generative attacks allow single-shot synthesis but often introduce perceptible artifacts or depend on computationally intensive architectures, while diffusion and autoregressive approaches incur high inference latency. To address this gap, we propose a generative attack framework operating in the continuous latent space of a neural audio codec. A conditional generator synthesizes class-specific perturbations in a single forward pass and decodes them into adversarial waveforms. Our method achieves targeted attack success rates up to $99\%$ with sub-$7$\,ms inference, outperforming generative baselines while reducing latency by $24\times$. 
\end{abstract}

\section{Introduction}
\label{sec:intro}

Advances in speech and audio processing have enabled the large-scale deployment of intelligent audio systems, ranging from automatic speaker verification (ASV) and biometric authentication~\cite{li2017deep, xiang2019margin, Desplanques2020ECAPATDNN} to environmental sound classification~\cite{tokozume2017learning}, acoustic scene analysis~\cite{barchiesi2015acoustic, ding2024acoustic}, and sound event detection~\cite{cakir2015polyphonic}. These technologies are now deeply embedded in consumer devices and services powered by deep neural networks (DNNs), including smart speakers (e.g., Apple HomePod, Amazon Echo) and voice assistants (e.g., Siri, Google Assistant)~\cite{edu2020smart}. While DNNs have driven substantial performance gains, their widespread adoption in real-world and safety-critical scenarios necessitates rigorous evaluation of potential security risks prior to deployment~\cite{lutz2021privacy, schuller2016interspeech}. 

A primary concern is the vulnerability of DNN-based audio models to adversarial attacks~\cite{goodfellow2014explaining, vakhshiteh2021adversarial}. Traditional audio adversarial attacks operate directly in the high-dimensional input representation space, manipulating either raw waveforms or time--frequency representations such as spectrograms. Methods including FGSM~\cite{goodfellow2014explaining}, PGD~\cite{madry2017towards}, and CW~\cite{carlini2018audio} optimize $L_p$-bounded perturbations via iterative gradient-based updates. While highly effective, their reliance on multiple forward and backward passes per sample introduces prohibitive computational overhead, rendering them largely \emph{impractical for real-time or streaming scenarios}, which are the most critical attack vectors for live user services~\cite{esmaeilpour2019robust, du2020sirenattack, du2020unified}.

Recent generative approaches, including the Fast Audio Perturbation Generator (FAPG)~\cite{xie2021enabling} and conditional GAN-based methods (CGAN)~\cite{wang2021fast}, improved efficiency via single-shot generation but remain confined to the high-dimensional waveform space, complicating optimization and often introducing perceptible artifacts. Inspired by Unrestricted Adversarial Examples (UAEs)~\cite{brown2018unrestricted} in vision, recent work has explored latent-space manipulation. However, most generative audio attacks—including VAE, diffusion, and autoregressive methods~\cite{qu2022synthesising, wang2024diffusion, ziv2025breaking}—depend on heavy task-specific architectures designed for automatic speech recognition, resulting in substantial inference overhead and limiting real-time applicability to audio classification and biometric verification.

Our work \textbf{addresses a critical gap}: evaluating real-time, highly efficient generative attacks against audio classification and speaker verification systems. Rather than relying on task-specific feature extractors or computationally intensive sampling, we introduce a universal, end-to-end differentiable adversarial framework operating in the continuous latent space of a general-purpose neural audio codec (Descript Audio Codec~(DAC)~\cite{kumar2023high}). Leveraging a pre-trained compressed acoustic latent manifold, our method generates high-quality adversarial perturbations in a single forward pass. This provides a key advantage: combining the stealth and structural fidelity of latent-space manipulation with \emph{ultra-low latency}, making it well suited for compromising real-time streaming audio systems where existing methods fall short.

Our key \textbf{contributions} are summarized as follows:
\begin{enumerate}
    \item \textbf{Latent-Space Generative Attack:} We propose a trainable conditional generator that operates directly in the continuous latent space of DAC\footnote{Code is available at~\href{https://github.com/vcbsl/DAC-GAN}{VCBSL/DAC-GAN}.}, enabling adversarial manipulation of high-level compressed acoustic representations rather than raw waveform perturbations.
    
    \item \textbf{Fully Differentiable Codec Pipeline:} We design an end-to-end differentiable framework in the codec’s pre-quantization space, overcoming the non-differentiability of discrete codebooks and enabling direct gradient-based optimization of latent perturbations.

    \item \textbf{Real-Time Generation:} A single forward pass generates adversarial examples in under $7$\,ms per sample, yielding up to $24\times$ speedup over generative baselines and $18{,}900\times$ over iterative methods such as C\&W.

    \item \textbf{High and Consistent Attack Success:} Our method achieves targeted Attack Success Rate (ASR) above 77\% (up to 99\%) across CNN- and transformer-based audio classification models, outperforming generative baselines and demonstrating strong architectural generality and real-time threat viability.
    
\end{enumerate}

\section{Proposed Method}
\label{sec:method}

We propose a white-box generative adversarial framework operating entirely in the continuous latent space of a neural audio codec. As illustrated in Figure~\ref{fig:pipeline}, the end-to-end differentiable generative pipeline comprises four components: (1) a frozen codec encoder mapping audio to latents; (2) a trainable conditional generator $G_\theta$; (3) a frozen codec decoder reconstructing waveforms; and (4) a fully differentiable acoustic preprocessor bridging the waveform to the frozen victim model $f_v$.

\begin{figure}[ht!]
    \centering
    \includegraphics[width=0.40\textwidth]{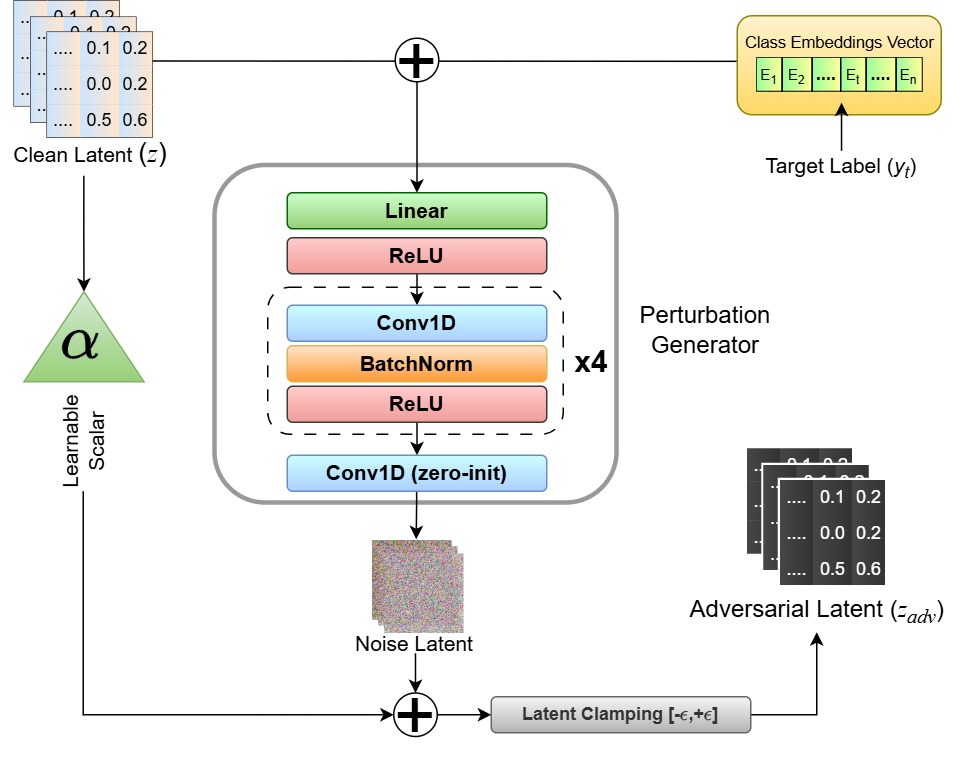}
    \caption{Detailed architecture of the Conditional Latent Generator ($G_\theta$). The network utilizes a multi-scale temporal receptive field (Conv1D) to process the fused continuous representation. A residual skip connection scaled by a learnable parameter $\alpha$ ensures training stability, while zero-initialization of the final layer guarantees negligible initial perturbation.}
    \label{fig:generator}
\end{figure}

\subsection{Latent Projection and Conditional Generation}

Let $x \in \mathbb{R}^{T}$ denote the raw audio waveform. The frozen Descript Audio Codec (DAC) encoder extracts a compressed, continuous latent representation. We adopt DAC over alternative representations~\cite{defossez2022high, zeghidour2021soundstream} due to its state-of-the-art compression and continuous latent manifold, which preserves acoustic fidelity and facilitates stable optimization: 

\begin{equation}
    z = E_{DAC}(x), \quad z \in \mathbb{R}^{C \times L}
\end{equation}
where $C$ is the channel dimension and $L \ll T$. This structured manifold allows minimal displacements to induce maximal semantic shifts without introducing jagged, audible artifacts.

To achieve single-shot generation, we introduce a trainable feed-forward generator $G_\theta$ (Figure~\ref{fig:generator}). It is conditioned on a target $t$, which can be a discrete class label $y_t$ (for classification) or a continuous speaker embedding $v_{tgt} \in \mathbb{R}^{D_{emb}}$ (for speaker verification). For ASV attack, the embedding is linearly projected to match the latent sequence dimensions of the DAC, yielding an embedding $E_{tgt} \in \mathbb{R}^{C \times L}$. 

As detailed in Figure~\ref{fig:generator}, the fused input ($z + y_t$) is processed through a Linear layer, a ReLU activation, and four stacked Conv1D-BatchNorm-ReLU blocks with decreasing kernel sizes to capture temporal dependencies. Rather than predicting the absolute latent, $G_\theta$ outputs a residual perturbation. A learned scaling parameter $\alpha$ stabilizes the skip connection, and the terminal Conv1D layer is zero-initialized to ensure gradual addition of noise. The bounded adversarial latent is clamped to prevent gradient explosion and is formulated as follows:
\begin{equation}
    \delta_z = \text{Clip}\big(G_\theta(z + E_{tgt}) + \alpha \cdot z, -\epsilon_{bnd}, \epsilon_{bnd}\big)
\end{equation}
\begin{equation}
    z_{adv} = \text{Clip}(z + \delta_z, -z_{max}, z_{max})
\end{equation}

\subsection{Differentiable Decoding and Feature Extraction}
The bounded latent sequence is passed through the frozen DAC decoder to reconstruct the adversarial waveform: \(x_{adv} = D_{DAC}(z_{adv})\). Because modern audio classifiers primarily operate on frequency-domain features, we formulate a differentiable preprocessor $\mathcal{P}$ that performs a Short-Time Fourier Transform (STFT), Mel-filterbank mapping, and logarithmic compression while preserving the computational graph. The resulting spectral feature map bridging the codec to the classifier is:
\begin{equation}
    S_{adv} = \mathcal{P}(x_{adv}), \quad S_{adv} \in \mathbb{R}^{F_{mel} \times T_{frames}}
\end{equation}

\subsection{Adversarial Objective Formulation}
To train the adversarial generator end-to-end without requiring iterative inference-time optimization, we define a unified composite loss function that accommodates both closed-set classification  (e.g., speech command or environmental sound classification) and open-set metric learning (e.g., speaker verification). The overall objective is formulated as:

\begin{equation}
    \mathcal{L}_{total} = \mathcal{L}_{adv} + \lambda_{m} \mathcal{L}_{margin}^k + \lambda_{L2} \frac{\lVert \delta_z \rVert_2}{B}
\end{equation}
where $B$ is the batch size, and the $L_2$ regularization term explicitly bounds the continuous perturbation $\delta_z$ within the DAC latent space to preserve acoustic fidelity. The task-specific components, $\mathcal{L}_{adv}$ and $\mathcal{L}_{margin}^k$ (where $k \in \{cls, emb\}$), are defined according to the victim model's operational domain:

\vspace{1mm}
\noindent \textbf{Discrete Categorical Objective:} For standard classifiers where the target prediction $\hat{y} = f_v(S_{adv})$ represents pre-softmax logits, the primary adversarial loss is the Cross-Entropy formulation, $\mathcal{L}_{adv} = \mathcal{L}_{CE}(\hat{y}, y_t)$. Because relying solely on cross-entropy can lead to vanishing gradients once the target class becomes marginally dominant, we explicitly widen the decision boundary to ensure high-confidence misclassification using a Carlini-Wagner margin loss~\cite{carlini2018audio}:
\begin{equation}
    \mathcal{L}_{margin}^{cls} = \max\Big(0, \max_{i \neq y_t} (\hat{y}_i) - \hat{y}_{y_t} + m_{cls}\Big)
\end{equation}
where $m_{cls}$ is a constant scalar margin enforcing strict separation between the target logit and the next most likely class.

\vspace{1mm}
\noindent \textbf{Continuous Embedding Objective:} For biometric verification systems (e.g., ECAPA-TDNN) operating in an open-set manifold, the victim model outputs a continuous representation vector $v_{adv} = f_v(S_{adv}) \in \mathbb{R}^{D_{emb}}$. Optimization in this space relies on geometric distances rather than discrete logits. Denoting the cosine similarity as $\cos(\cdot,\cdot)$, the adversarial loss maximizes alignment with the target speaker embedding $v_{tgt}$ via $\mathcal{L}_{adv} = 1 - \cos(v_{adv}, v_{tgt})$. To ensure the adversarial embedding is simultaneously separated from the original source speaker $v_{src}$, the margin loss is formulated geometrically:
\begin{equation}
    \mathcal{L}_{margin}^{emb} = \max\Big(0, \cos(v_{adv}, v_{src}) - \cos(v_{adv}, v_{tgt}) + m_{emb}\Big)
\end{equation}
where $m_{emb}$ dictates the requisite separation threshold in the cosine space, ensuring that the adversarial embedding robustly shifts past the verification system's decision threshold.


To stabilize the non-convex adversarial optimization landscape and reduce batch-level gradient volatility, $G_\theta$'s weights are maintained using an Exponential Moving Average (EMA). During training, active weights $\theta$ are updated via the optimizer, while a shadow copy is decayed by a factor $\beta$: $\theta_{EMA} \leftarrow \beta \theta_{EMA} + (1 - \beta) \theta$. Inference exclusively utilizes $\theta_{EMA}$, yielding smooth, highly transferable adversarial perturbations.

\vspace{-2mm}

\section{Experimental Protocol}
\label{sec:experiments}

\noindent\textbf{Datasets}: We evaluate our adversarial attack generator on four benchmark datasets spanning speech commands~\cite{warden2018speech}, acoustic scene classification~\cite{mesaros2018multi}, environmental sound classification~\cite{Salamon:UrbanSound:ACMMM:14}, and speaker verification~\cite{panayotov2015librispeech}. Google Speech Commands~\cite{warden2018speech} contains $35$ one-second spoken words at $16$ kHz (80,000 train / 4,273 test). TAU Urban Acoustic Scenes 2019~\cite{mesaros2018multi} comprises 10 urban scene classes with 10-second clips (7,348 train / 1,837 val / 4,185 test). UrbanSound8k~\cite{Salamon:UrbanSound:ACMMM:14} includes 10 environmental sound classes across 10 predefined folds; we use Folds 1--3 and 6--10 for training (6,806 files), Fold 4 for validation (990), and Fold 5 for testing (936). For speaker verification, LibriSpeech (train-clean-100)~\cite{panayotov2015librispeech} provides 251 speakers; selecting 3 reference and 5 test utterances per speaker yields 1,255 genuine and 313,750 imposter trials.

\vspace{0.5mm}

\noindent\textbf{Baselines and evaluation metrics:}
All gradient-based baselines were implemented with consistent hyperparameters for fair comparison. FGSM and PGD used a perturbation bound of $\epsilon=0.01$, while PGD used a step size $\alpha=0.01$ for 10 iterations. The C\&W attack used $c_{const}=1$, a learning rate of $0.01$, and 50 optimization iterations. For generative baselines, we implemented FAPG~\cite{xie2021enabling} with a Wave-U-Net architecture~\cite{stoller2018wave}, and CGAN~\cite{wang2021fast} with a conditional generator–discriminator trained using WGAN-GP~\cite{adler2018banach} to generate targeted waveform-level perturbations. Both FAPG and CGAN are representative single-shot generative attacks for fast inference.

For the victim models, we use: (1) \textit{Audio Spectrogram Transformer}~\cite{gong2021ast} for Speech Commands obtaining a 98.37\% accuracy; (2) \textit{PANNs CNN14}~\cite{kong2020panns} fine-tuned for acoustic scene classification; (3) \textit{PANNs CNN14} fine-tuned on TAU Urban Acoustic Scenes $2019$ and UrbanSound8k by unfreezing its 5th and 6th convolution and fully connected layers and classifier head. Achieves an accuracy of 76.65\% and 86.22\%, on test set of TAU Urban Acoustic Scenes 2019 and UrbanSound8k, respectively. (4) \textit{ECAPA-TDNN}~\cite{Desplanques2020ECAPATDNN} with cosine similarity used for Speaker Verification obtaining 99.66\% accuracy and 0.33\% EER on clean Genuine and Imposter Trials of LibriSpeech. We evaluate attack efficacy using two \textbf{metrics}: \textit{Attack Success Rate (ASR)} and model \textit{Accuracy (Acc)}. In untargeted setting, ASR measures the percentage of adversarial samples misclassified into any incorrect class. In the targeted setting, ASR quantifies the proportion of samples classified as the specified target class. Acc denotes the percentage of samples that retain their original label. Importantly, in targeted attacks, ASR and Accuracy are not direct inverses: a low Accuracy with a high (but imperfect) targeted ASR indicates that some samples failed to reach the exact target yet were still pushed away from the true label, effectively succeeding as untargeted attacks.

\vspace{0.5mm}

\noindent\textbf{Implementation Details:} We use the 16kHz DAC frozen encoder--decoder (compression $\sim$$320\times$) model for mapping waveforms to continuous latents of dimension $C \times L$ ($L \ll T$) because of its state-of-the-art neural audio tokenizer performance enabling \textit{high-quality, low-bitrate audio}. The conditional generator operates directly in this latent space using stacked Conv1D--BN--ReLU blocks (1024$\rightarrow$1024$\rightarrow$512$\rightarrow$256) followed by a $1\times1$ convolution to predict a residual perturbation. The latent offset is clamped to $[-2,2]$, and final latents to $[-5,5]$ before decoding, to prevent gradient explosion. Experiments were conducted on NVIDIA RTX 5000 Ada and A10 GPUs with CUDA.

\section{Results and Discussion}

\vspace{-2mm}

\begin{table}[ht!]
\centering
\small
\renewcommand{\arraystretch}{0.9}
\setlength{\tabcolsep}{4pt}
\begin{tabular}{
l
S[table-format=2.2]
S[table-format=2.2]
S[table-format=2.2]
S[table-format=2.2]
S[table-format=2.4, round-mode=none]
}
\toprule
& \multicolumn{2}{c}{\textbf{Untargeted (\%)}} 
& \multicolumn{2}{c}{\textbf{Targeted (\%)}} 
& \textbf{Time (sec)} \\
\cmidrule(lr){2-3} \cmidrule(lr){4-5}
\textbf{Method} 
& {Acc} & {ASR} 
& {Acc} & {ASR} 
& {(1 sample)} \\
\midrule
FGSM & 91.12 & 8.88 & 93.46 & 3.63 & 0.9511 \\
PGD  & 54.35 & 45.65 & 85.81 & 12.63 & 2.4880 \\
CW   & 22.40 & 77.60 & 32.69 & 66.22 & 13.2731 \\
FAPG &  18.92  &  82.08 & 3.53  & 80.77   & 0.0153 \\
CGAN & 5.15  & 93.72 & 6.42  & \textbf{93.56} & 0.0158  \\
\textbf{Ours} & \textbf{3.42} & \textbf{96.58} & \textbf{3.32} & 77.65 & \textbf{0.0067} \\
\bottomrule
\end{tabular}
\caption{Performance Comparison under Untargeted and Targeted Settings on Speech Commands dataset.}
\label{tab:speech_commands_results}
\end{table}

\vspace{-7mm}

Table~\ref{tab:speech_commands_results} reports results on the \textit{Google Speech Commands} dataset evaluated with AST (98.37\% clean accuracy). Our latent-space attack achieves the strongest overall performance. In the untargeted setting, we obtain \textbf{96.58\% ASR}, outperforming gradient-based FGSM (+87.70\%), PGD (+50.93\%), and C\&W (+18.98\%), while also exceeding FAPG (+14.50\%) and CGAN (+2.86\%). In the targeted setting, although CGAN attains the highest ASR (93.56\%), our method remains competitive (77.65\%) while significantly surpassing FGSM (+74.02\%), PGD (+65.02\%), C\&W (+11.43\%), and FAPG (-3.12\% lower). This variance suggests that manipulating localized phonemes in 1-second clips via global latents is harder than shifting continuous scenes or speaker identities. Importantly, our \emph{approach achieves the lowest latency} at \textbf{0.0067\,s} per sample, making it faster than both gradient-based methods (up to 1,980$\times$ over C\&W) and prior generative attacks, demonstrating superior efficiency with strong attack effectiveness.

\vspace{-1mm}

\begin{table}[ht!]
\centering
\small
\renewcommand{\arraystretch}{0.9}
\setlength{\tabcolsep}{4pt}
\begin{tabular}{
l
S[table-format=2.2]
S[table-format=2.2]
S[table-format=2.2]
S[table-format=2.2]
S[table-format=2.4, round-mode=none]
}
\toprule
& \multicolumn{2}{c}{\textbf{Untargeted (\%)}} 
& \multicolumn{2}{c}{\textbf{Targeted (\%)}} 
& \textbf{Time (sec)} \\
\cmidrule(lr){2-3} \cmidrule(lr){4-5}
\textbf{Method} 
& {Acc} & {ASR} 
& {Acc} & {ASR} 
& {(1 sample)} \\
\midrule
FGSM & 48.5042 & 44.8574 & 60.9401 & 12.1581 & 0.7140 \\
PGD  & 43.0555 & 50.0619 & 46.5491 & 30.2136 & 1.2400 \\
CW   &  4.7008 & 94.6716 & 14.3696 & 92.2057 & 6.1400 \\
FAPG &  2.10  & 97.90  & 1.76  &  96.52  &  0.0048   \\
CGAN & 20.17 & 79.83 & 21.22 & 77.13 &  0.0237 \\
\textbf{Ours} &  \textbf{0.89}   &  \textbf{99.11}   &  \textbf{1.23}   &  \textbf{97.17}   & \textbf{0.0039} \\
\bottomrule
\end{tabular}
\caption{Performance Comparison under Untargeted and Targeted Settings for Environmental Sound Classification Task on UrbanSound8K dataset.}
\label{tab:attack_results_urbansound8k}
\end{table}

\vspace{-10mm}

\begin{table}[ht!]
\centering
\small
\renewcommand{\arraystretch}{0.9}
\setlength{\tabcolsep}{4pt}
\begin{tabular}{
l
S[table-format=2.2]
S[table-format=2.2]
S[table-format=2.2]
S[table-format=2.2]
S[table-format=2.4, round-mode=none]
}
\toprule
& \multicolumn{2}{c}{\textbf{Untargeted (\%)}} 
& \multicolumn{2}{c}{\textbf{Targeted (\%)}} 
& \textbf{Time (sec)} \\
\cmidrule(lr){2-3} \cmidrule(lr){4-5}
\textbf{Method} 
& {Acc} & {ASR} 
& {Acc} & {ASR} 
& {(1 sample)} \\
\midrule
FGSM & 13.3094 & 84.0399 & 15.2700 & 14.6700 & 1.8900 \\
PGD  &  6.5900 & 92.2100 & 11.9115 & 77.2735 & 6.5500 \\
CW   &  0.3106 & 99.8441 & \textbf{0.24}  & \textbf{96.65}   & 38.5500 \\
FAPG & 2.95  & 97.02 &  4.34 &  95.80  &  0.0097   \\
CGAN & 1.2 & 98.25 & 4.12 & 95.97 &  0.0232 \\
\textbf{Ours} &  \textbf{0}   &  \textbf{100}   &  0.32   &  94.07   & \textbf{0.0056} \\
\bottomrule
\end{tabular}
\caption{Performance Comparison under Untargeted and Targeted Settings for Environmental Sound Classification Task on DCASE2019 dataset.}
\label{tab:attack_results}
\end{table}

\vspace{-5mm}

For environmental sound classification on \textit{UrbanSound8K} and \textit{DCASE2019} (Tables~\ref{tab:attack_results_urbansound8k} and \ref{tab:attack_results}), evaluated using PANNs CNN14, our method consistently achieves the strongest overall attack performance with the lowest latency. On UrbanSound8K (86.22\% clean accuracy), we obtain \textbf{99.11\%} untargeted ASR, surpassing FGSM (+54.25\%), PGD (+49.05\%), C\&W (+4.44\%), and further improving over generative baselines FAPG (+1.21\%) and CGAN (+19.28\%). In the targeted setting, our \textbf{97.17\%} ASR exceeds FGSM (+85.01\%), PGD (+66.96\%), C\&W (+4.96\%), FAPG (+0.65\%), and CGAN (+20.04\%).

On DCASE2019 (76.65\% clean accuracy), we achieve \textbf{100\%} untargeted ASR, outperforming FGSM (+15.96\%), PGD (+7.79\%), C\&W (+0.16\%), FAPG (+2.98\%), and CGAN (+1.75\%). In the targeted case, C\&W attains the highest ASR (96.65\%), while our method achieves a competitive \textbf{94.07\%}, exceeding FGSM (+79.40\%), PGD (+16.80\%), and remaining close to FAPG (-2.45\%) and CGAN (-1.90\%). Importantly, our inference time (\textbf{0.0039--0.0056\,s}) is the lowest across all methods, making it substantially faster than both iterative optimization and prior generative attacks, thereby demonstrating superior efficiency \textit{with near-complete model collapse} across environmental sound domains.

\vspace{-3mm}

\begin{table}[ht!]
\centering
\small
\renewcommand{\arraystretch}{0.9}
\setlength{\tabcolsep}{4pt}
\begin{tabular}{
l
S[table-format=2.2]
S[table-format=2.2]
S[table-format=2.2]
S[table-format=2.2]
S[table-format=2.2]
}
\toprule
& \multicolumn{2}{c}{\textbf{Untargeted (\%)}} 
& \multicolumn{2}{c}{\textbf{Targeted (\%)}} 
& \textbf{Time (sec)} \\
\cmidrule(lr){2-3} \cmidrule(lr){4-5}
\textbf{Method} 
& {Acc} & {ASR} 
& {Acc} & {ASR} 
& {(1 sample)} \\
\midrule
FGSM & 90.68 & 9.32 & 97.92 & 2.06 & 0.08 \\
PGD  & 0.38  & 99.60 & 87.57 & 12.43 & 1.98 \\
CW   & 0.03  & 99.95 & \multicolumn{1}{c}{17.45} 
     & \multicolumn{1}{c}{82.64} & 66.2 \\
FAPG & 19.00 & 81.00 & 21.60 & 78.40 & 0.0836 \\
\textbf{Ours} & \multicolumn{1}{c}{\textbf{0.00}}
     & \multicolumn{1}{c}{\textbf{100}}
     & \multicolumn{1}{c}{\textbf{0.02}} 
     & \multicolumn{1}{c}{\textbf{99.80}} 
     & \multicolumn{1}{c}{\textbf{0.0035}} \\
\bottomrule
\end{tabular}
\caption{Performance Comparison under Untargeted and Targeted Settings for Speaker Verification on LibriSpeech dataset.}
\vspace{-4mm}
\label{tab:ecapa_librispeech_compact}
\end{table}

\vspace{-5mm}

For speaker verification on \textit{LibriSpeech} (Table~\ref{tab:ecapa_librispeech_compact}) evaluated against ECAPA-TDNN (99.66\% clean accuracy), our method achieves \textbf{100\%} untargeted ASR, improving over FGSM by \textbf{+90.68\%}, PGD by \textbf{+0.40\%}, C\&W by \textbf{+0.05\%}, and FAPG by \textbf{+19\%}. For targeted attacks, we obtain \textbf{99.80\% ASR}, exceeding FGSM by \textbf{+97.74\%}, PGD by \textbf{+87.37\%}, C\&W by \textbf{+17.16\%}, and FAPG by \textbf{+21.40\%}. Note that CGAN could not be evaluated due to mode collapse during training. 
Unlike waveform-domain optimization, which must iteratively push embeddings across cosine decision thresholds, our latent generator directly learns to induce decisive embedding displacement within the compact DAC manifold. Critically, this is achieved with only \textbf{0.0035\,s} latency per sample, making our approach \textbf{23.8$\times$} faster than the generative FAPG baseline, and between \textbf{18.6$\times$} and \textbf{18,914$\times$} faster than iterative methods.

\section{Conclusion}
\label{sec:conclusion}
The widespread deployment of streaming and on-device audio systems demands rigorous evaluation against real-time adversarial threats. Prior generative approaches reduce optimization cost but remain waveform-bound, task-specific, or lack targeted control. We introduce an end-to-end differentiable framework that generates adversarial examples directly in the latent space of a neural audio codec, eliminating iterative optimization through a single forward pass. The method achieves targeted success rates of \textbf{97.17\%} on UrbanSound8K and \textbf{77.65\%} on Speech Commands in under \textbf{7 ms} per sample, substantially faster than strong gradient-based baselines. These findings identify compressed semantic latent manifolds as a powerful yet unexplored adversarial surface, enabling both high attack success and real-time generation. Future work will investigate black-box transferability and latent-space defenses.


\section{Generative AI Use Disclosure}
Generative AI tools were used solely for language refinement, clarity improvement, and minor formatting adjustments. The research conception, methodology, experimental design, implementation, results, and analysis were entirely conducted by the authors. All AI-assisted edits were carefully reviewed, validated, and approved by the authors, who take full responsibility for the final content of this manuscript.

\bibliographystyle{IEEEtran}
\bibliography{mybib}

\end{document}